\documentclass[twocolumn, a4paper]{revtex4}

\usepackage[T1]{fontenc}
\usepackage{mathrsfs}
\usepackage{amsmath}
\usepackage{amssymb}
\usepackage{graphicx}
\usepackage{braket}
\usepackage{physics}
\usepackage{bm}
\usepackage{bbold}
\usepackage{mathtools}
\usepackage{color}
\usepackage[version=3]{mhchem}

\begin{document}
	
\title{ Valley-dependent electron optics using quantum dots in bilayer graphene}

\author{Fereshte Ildarabadi}
\affiliation{School of Physical Sciences, Dublin City University, Glasnevin, Dublin 9, Ireland}
\author{Stephen R. Power}
\email{stephen.r.power@dcu.ie}
\affiliation{School of Physical Sciences, Dublin City University, Glasnevin, Dublin 9, Ireland}


\begin{abstract}
Electrostatically defined quantum dots (QDs) with layer-antisymmetric gating in Bernal-stacked bilayer graphene (BLG) open a local gap and generate a mass-like term with opposite sign in the two valleys, producing strongly valley-dependent scattering without magnetic fields, strain, or spin–orbit coupling. Building on this mechanism, we propose a tunable platform based on such QDs for valley-dependent electron optics in BLG. Using a four-band continuum model and a generalized multiple-scattering formalism, we analyze the scattering of Gaussian electron beams from single- and multi-dot architectures and compute valley-resolved currents and angular profiles. A single dot produces distinct valley-dependent deflection, while multi-dot configurations enable enhanced control: identical-dot arrays act as valley splitters, whereas oppositely gated pairs function as valley filters. Combining these elements yields tunable generation, steering, and filtering of highly valley-polarized currents with strong suppression of forward transmission. The required energy scales, gate asymmetries, and device dimensions are within experimentally accessible regimes for dual-gated BLG, establishing quantum-dot arrays as a realistic platform for controllable valley-resolved electron optics.

\end{abstract}

\maketitle

\section{Introduction}
The valley degree of freedom in two-dimensional materials has attracted growing interest as a potential carrier of information, complementing charge and spin in next-generation electronic platforms. In materials with multiple inequivalent extrema in momentum space, electrons can occupy distinct valleys, forming the basis of valleytronics~\cite{rycerz2007valley,schaibley2016valleytronics,vitale2018valleytronics,xu2025valleytronics,seyler2026valleytronics}. Graphene and its multilayer counterparts are prototypical systems in this context, hosting two inequivalent valleys, $K$ and $K'$, at the corners of the Brillouin zone that can serve as a binary degree of freedom for information encoding. Most early valleytronics work focused on the generation and detection of valley  polarization~\cite{rycerz2007valley,xiao2012coupled,sui2015gate,li2020room,shimazaki2015generation}; building functional devices, however, requires the ability to transport, steer, and filter valley-resolved currents in a controlled way. This motivates extending the framework of electron optics in graphene to the valley degree of freedom~\cite{hawkes2022principles,yu2022electron,belayadi2024valley,nguyen2016valley,li2018valley,garcia2008fully}.

Electron optics in graphene exploits the chiral, ballistic nature of its low-energy carriers~\cite{katsnelson2006chiral,ildarabadi2025tunable,chen2016electron,bai2018generating,elahi2019impact,paredes2021gradient,liu2017creating} to manipulate electron beams in close analogy with conventional optics. This leads to phenomena such as Klein tunneling through electrostatic barriers~\cite{katsnelson2006chiral}, Veselago-type negative refraction and lensing at $p$-$n$ junctions~\cite{cheianov2007focusing,paredes2021gradient}, electron collimation, and highly directional ballistic transport. More elaborate architectures, including Fabry-P\'erot interferometers~\cite{shytov2008klein,young2009quantum}, electron waveguides~\cite{forrester2023electron,liang2001fabry}, multi-terminal beam-splitter geometries~\cite{lima201650,brandimarte2017tunable}, and electrostatically defined quantum-dot scatterers~\cite{zhao2015creating,jiang2017tuning,heinisch2013mie,lee2016imaging}, further expand the range of electron-optic functionalities. In particular, circular gated dots act as compact scatterers supporting Mie-like resonances and strongly anisotropic angular profiles, and can be combined into multi-dot geometries for beam shaping, splitting, and collimation \cite{boggild2017two,zhao2023electron,wan2021dirac}. However, in the absence of additional ingredients, such structures affect the $K$ and $K'$ valleys identically and therefore lack intrinsic valley selectivity.

Achieving valley selectivity requires breaking the $K\leftrightarrow K'$ symmetry. Several mechanisms have been explored, including strain-induced pseudomagnetic fields~\cite{hsu2020nanoscale,yu2022electron,banerjee2020strain}, the valley Hall effect in gapped systems driven by Berry-curvature-induced transverse responses~\cite{yin2022tunable}, and engineered structural inhomogeneities or curvature-induced deformations~\cite{pan2025topological,carrillo2018enhanced,stegmann2016current,settnes2016graphene}. While these approaches demonstrate valley-contrasting behavior, they typically rely on fixed structural features or externally applied strain, which limits the in-situ tunability required for electron-optic applications.

Electrostatic gating provides a more flexible route to valley control~\cite{sui2015gate,chen2020gate}. In monolayer graphene, a sublattice-asymmetric potential acts as a mass term, opens a band gap, and induces valley-contrasting transport~\cite{aktor2021valley,ando2015theory,gorbachev2014detecting,hunt2013massive}, but realizing such control experimentally remains challenging. Bilayer graphene (BLG) offers a more practical alternative. In Bernal-stacked BLG, a perpendicular electric field applied via top and bottom gates breaks inversion symmetry between the layers, opens a tunable band gap~\cite{solomon2021valley,zhang2009direct}, and generates valley-contrasting Berry curvature. Reversing the gate polarity flips the sign of the response at each valley~\cite{solomon2021valley,sui2015gate,yin2022tunable}, enabling direct electrical control of valley-resolved transport. This coupling between layer and valley degrees of freedom makes BLG a suitable platform for tunable valleytronic functionalities.

\begin{figure}
    \centering
    \includegraphics[width=0.9\linewidth]{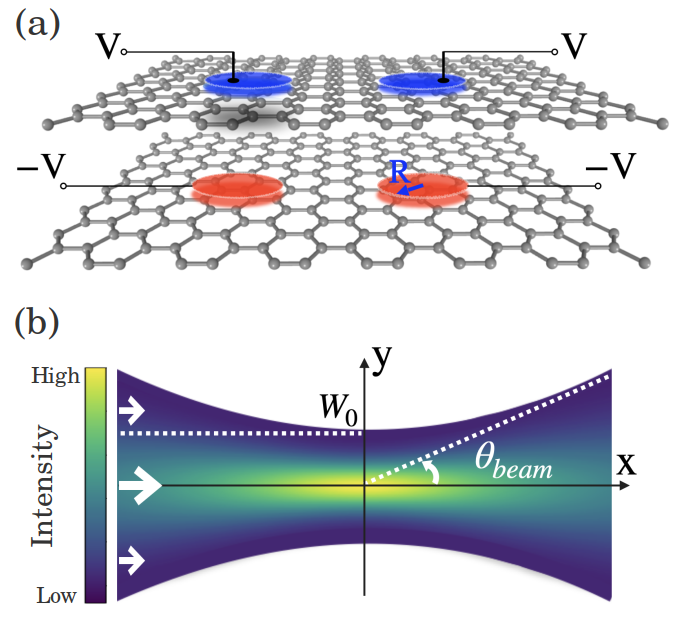}
\caption{(a) Schematic of two oppositely gated QDs with radius $R$ in BLG. (b) Gaussian beam propagating along the $x$ direction. $W_0$ is the beam waist, and $\theta_{{beam}}$ denotes the characteristic divergence angle.}
    \label{fig:fig0}
\end{figure}

The combination of electron-optic control, electrostatically defined quantum dots (QDs), and the layer–valley coupling in BLG naturally suggests that QDs with layer-antisymmetric gating can act as valley-contrasting scatterers controlled purely by electrostatics. However, a systematic framework for valley-dependent electron optics based on such structures is still lacking. 
In this work, we introduce such QDs as building blocks for valley-dependent electron optics in BLG, as illustrated schematically in Fig.~\ref{fig:fig0}(a). This gating generates a spatially localized mass-like term that couples with opposite sign to the two valleys, producing strongly valley-resolved scattering without requiring magnetic fields or strain. The problem is treated within a four-band Dirac model combined with a generalized Mie expansion, with Gaussian incident electron beams (as shown in Fig.~\ref{fig:fig0} (b)) incorporated via projection onto partial-wave modes. The scattered current develops tunable distinct angular profiles for the two valleys, enabling angular separation, valley splitting, valley filtering and valley-selective beam steering providing a scalable and experimentally feasible route to controlling valley-resolved currents.

We begin in Sec.~\ref{method} by outlining the general theoretical framework for scattering of an electronic plane-wave from a QD in BLG, introducing the relevant physical quantities and briefly summarizing key results from previous work in Sec.~\ref{Current}. We then extend the analysis to Gaussian electron beams for a single dot and compare the resulting current distributions with the plane-wave case in Sec.~\ref{Gaussian}, followed by a deeper analysis of the resulting current decomposition in Sec.~\ref{decomposition}. The multiple-scattering formalism for electrons in BLG is subsequently developed and applied to multi-dot architectures in Sec.~\ref{multiple scattering}. The corresponding results, which constitute the main findings of this work, are presented in Sec.~\ref{Results}, including identical-dot, oppositely gated, and multicomponent configurations. Finally, possible experimental implementations are discussed before the conclusions in Sec.~\ref{Conclusion}.

\section{Method}\label{method}

We model electron scattering in AB-stacked BLG using the four-band continuum Hamiltonian in the basis $(A_2, b_2, a_1, B_1)$,where $A_2$ and $b_2$ label the sublattices of the upper layer and $a_1$ and $B_1$ those of the lower layer. $A_2$ lies directly above $B_1$, forming the interlayer dimer pair with coupling $\gamma_1$. Near the $K$ and $K'$ valleys ($\tau=\pm1$), the Hamiltonian in polar coordinates reads \cite{solomon2021valley}

\begin{equation} \label{Hamiltonian}
H(\mathbf{k})=\hbar v_F
\begin{pmatrix}
\tilde V_2 & \tau k e^{-i\tau\theta} & 0 & \tilde\gamma_1 \\
\tau k e^{i\tau\theta} & \tilde V_2 & 0 & 0 \\
0 & 0 & \tilde V_1 & \tau k e^{-i\tau\theta} \\
\tilde\gamma_1 & 0 & \tau k e^{i\tau\theta} & \tilde V_1
\end{pmatrix},
\end{equation}
where $k=|\mathbf{k}|$, $\theta=\arg\mathbf{k}$ and $\tilde X = X/(\hbar v_F)$.


Inside the dot, the interlayer asymmetry is defined as $\Delta = V_1 - V_2$, and the allowed wavevectors are
\begin{equation}
k^{\pm}
=
\sqrt{
\tilde{\varepsilon}_c^{\,2}
+
\left(\frac{\tilde{\Delta}}{2}\right)^2
\;\mp\;
\tilde{\gamma}_1
\sqrt{
\tilde{\varepsilon}_c^{\,2}
\left(1+\frac{\tilde{\Delta}^{\,2}}{\tilde{\gamma}_1^{\,2}}\right)
-
\left(\frac{\tilde{\Delta}}{2}\right)^2
}
},
\end{equation}
where $\tilde{\varepsilon}_c = E - (V_1 + V_2)/2$ is the energy measured from the band center. $k^{-}$ and $k^{+}$ correspond, respectively, to the lower ($\eta_2=-1$) and higher ($\eta_2=+1$) energy bands.

Outside the dot ($V_1 = V_2 = 0$), the allowed wavevectors reduce to
\begin{equation}
k^{\pm}_{0} = \sqrt{|\tilde{E}|\left(|\tilde{E}| \mp \tilde{\gamma}_1\right)}.
\end{equation}

The circular symmetry of the dot makes the total angular momentum projection a conserved quantum number, so the scattering problem decouples into independent channels labeled by $m \in \mathbb{Z}$. In each channel, the BLG spinor modes outside the dot, built from regular Bessel functions, take 
the form
\begin{equation}\label{psi_0}
\psi^{0,J}_{m}(kr,\theta)=
\begin{pmatrix}
|\tilde{E}|\, J_m(kr)\,e^{im\theta}\\
i\eta_1 k\, J_{m+\tau}(kr)\,e^{i(m+\tau)\theta}\\
-i\eta_2 k\, J_{m-\tau}(kr)\,e^{i(m-\tau)\theta}\\
\eta_1\eta_2 |\tilde{E}|\, J_m(kr)\,e^{im\theta}
\end{pmatrix},
\end{equation}
where $\eta_1=\pm1$ labels conduction/valence bands.

The incident plane-wave propagating along $x$ is decomposed using the standard identity
$e^{ikx}=\sum_{m=-\infty}^{\infty} i^{m}\, J_m(kr)\, e^{im\theta}$,
from which the incident spinor in channel $m$ is constructed as
\begin{equation}
\psi_{\mathrm{inc}}^m(k,r)
= \frac{i^m}{\sqrt{2(\tilde{E}^2+|k|^2)}}\,\psi^{0,J}_m(kr,\theta),
\end{equation}
and the full incident wave is $\psi_{\mathrm{inc}} = \sum_m \psi_{\mathrm{inc}}^m$.

The scattered wave outside the dot is expressed using outgoing Hankel functions in Eq. \eqref{psi_0} as $\psi^{0,H}_m$,
\begin{equation}
\psi_{\mathrm{sc}}(k,r)
= \frac{1}{\sqrt{2(\tilde{E}^2+|k|^2)}}
\sum_{m=-\infty}^\infty
c^{\mathrm{sc}}_{m,k}\, i^m\, \psi^{0,H}_m(kr,\theta),
\end{equation}
and the transmitted wave inside the dot is
\begin{equation}
\psi^{\mathrm{tr}}(k,r)
=\sum_{m=-\infty}^\infty
c^{\mathrm{tr}}_{m,k}\, i^m\, \psi^{V}_{m}(kr,\theta),
\end{equation}
where 
\begin{equation}
\psi^{V,J}_{m}(kr,\theta)=
\begin{pmatrix}
\tilde{\gamma}_1(\tilde{E}-\tilde{V}_1)(\tilde{E}-\tilde{V}_2)\, J_m(kr)\,e^{im\theta}\\
i\tilde{\gamma}_1(\tilde{E}-\tilde{V}_1) k\, J_{m+\tau}(kr)\,e^{i(m+\tau)\theta}\\
-i[(\tilde{E}-\tilde{V}_2)^2-k^2] k\, J_{m-\tau}(kr)\,e^{i(m-\tau)\theta}\\
[(\tilde{E}-\tilde{V}_2)^2-k^2] (\tilde{E}-\tilde{V}_1)\, J_m(kr)\,e^{im\theta}
\end{pmatrix}.\nonumber
\end{equation} 

The coefficients $c^{\mathrm{sc}}_{m,k}$ and $c^{\mathrm{tr}}_{m,k}$ in the latter two equations correspond to the scattering and transmission amplitudes, respectively, and are determined by imposing continuity of the wave function at the dot boundary $r=R$ for all four spinor components:
\begin{align}\label{continuity}
\psi_{\mathrm{inc}}(k_-^0,R)
+\psi_{\mathrm{sc}}(k_-^0,R)
+\psi_{\mathrm{sc}}(k_+^0,R) \nonumber\\
=\psi_{\mathrm{tr}}(k_-,R)
+\psi_{\mathrm{tr}}(k_+,R).
\end{align}
Since $k_-$ corresponds to a propagating mode while $k_+$ is evanescent, only $k_-^0$ contributes to the incident wave. This leads to a linear system for the coefficients $c^{\mathrm{sc}}_{m,k}$ and $c^{\mathrm{tr}}_{m,k}$, which can be solved by eliminating a subset of variables and back-substituting into the remaining equations. The approach introduced in this section follows that of Ref.~\cite{solomon2021valley}, where further details can be found.

We note that in practice, the angular momentum sum is truncated at a finite cutoff $|m| \leq M$. The required value of $M$ depends on the electron energy, the spatial region of study and the number of dots that we focus on later in this work. At higher values of the mentioned parameters, higher-order modes contribute more significantly and a larger cutoff is needed. To ensure a single consistent truncation that is sufficient for all configurations considered in this work, we use $M = 50$ throughout, although this value is larger than  cutoff value required for convergence in many cases.

\subsection{Current and valley-resolved quantities}\label{Current}

For each valley $\tau $, the current density operator is obtained from $\mathbf{j} = \partial H / \partial \mathbf{k}$, giving
\begin{equation}
\mathbf{j}_\tau
= v_F\, \psi_\tau^\dagger
\left(I_2 \otimes \boldsymbol{\sigma}_\tau\right)
\psi_\tau,
\quad
\boldsymbol{\sigma}_\tau = (\tau\sigma_x, \sigma_y),\nonumber
\end{equation}
where $I_2$ is the identity matrix in layer space and $\boldsymbol{\sigma}$ denotes the usual Pauli matrices in sublattice space. The corresponding charge density is given by $\rho_\tau = \psi_\tau^\dagger \psi_\tau$.

In the far field regime, the evanescent branch $k_0^+$ decays exponentially and therefore does not contribute to the radiated current, such that only the propagating branch $k_0^-$ remains. Using this together with the asymptotic form of the Hankel functions, the radial component of the scattered current at far-field reduces to \cite{solomon2021valley}
\begin{equation}
j^{\tau,\mathrm{rad}}_{\mathrm{sc}}(\theta)
\propto
\sum_{m,n}
c^{\mathrm{sc}^*}_{m,-},
c^{\mathrm{sc}}_{n,-},
e^{i(n-m)\theta}.
\end{equation}
This quantity is commonly used to characterize the angular dependence of scattering in quantum dot systems \cite{solomon2021valley,aktor2021valley,sadrara2019dirac,heinisch2013mie}.

The valley selectivity of the resulting current distribution can be quantified through the valley polarization, defined at each point in the system as

\begin{equation} \label{polarization}
P(r,\theta)
=\frac{j^{K}(r,\theta) - j^{K'}(r,\theta)}
     {j^{K}(r,\theta) + j^{K'}(r,\theta)},
\end{equation}
with $P = \pm 1$ corresponding to complete polarization into the $K$ or $K'$ valley, respectively.

\begin{figure}
    \centering
    \includegraphics[width=1\linewidth]{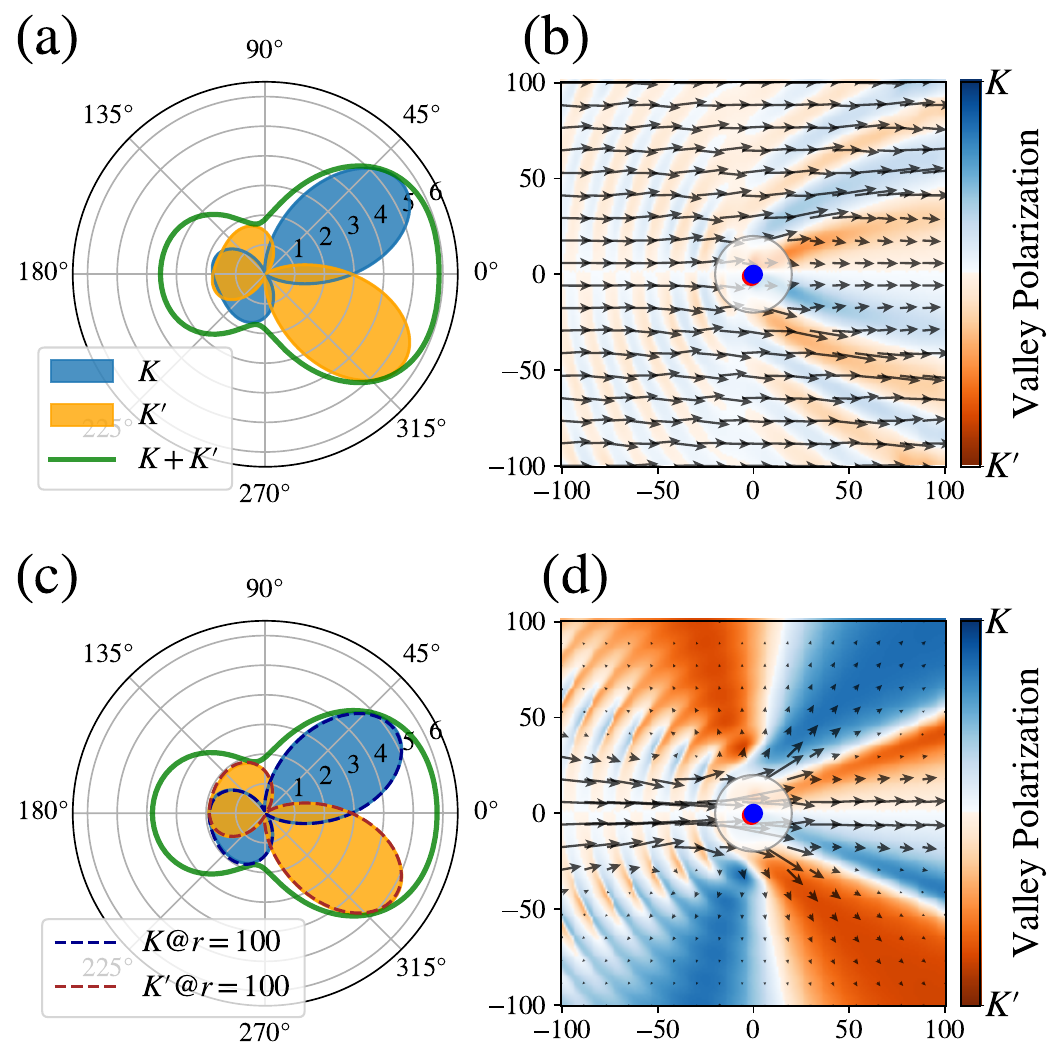}
    \caption{(a) Far-field angular dependence of the valley-resolved and total ($K+K'$) scattered current for plane-wave incidence; (b) Corresponding total current flow (arrows) and valley polarization (color map). (c) Same angular dependence for a Gaussian beam; dashed curves indicate currents evaluated at finite distance; (d) Corresponding current flow and valley polarization.}
    \label{fig:placeholder}
\end{figure}

Fig.~\ref{fig:placeholder}(a) and (b) illustrate these quantities for the scattering of a plane-wave with energy $E = 0.05\,\gamma_1$ from a single biased QD of radius $R = 5\,l_0$ with layer asymmetry $\Delta = -0.4\,\gamma_1$. Throughout this work we use $\gamma_1 = 0.38~\mathrm{eV}$ and $l_0 = \hbar v_F/\gamma_1 \approx 2~\mathrm{nm}$ as the units of energy and length, respectively.
Fig.~\ref{fig:placeholder}(a) shows the far-field angular dependence of the scattered current for the $K$ (blue) and $K'$ (orange) valleys, together with the combined total current (green). The two valleys exhibit clearly distinct angular profiles, demonstrating valley-dependent scattering from a biased QD. Fig.~\ref{fig:placeholder}(b) presents the corresponding total spatial current distribution where the length of arrows indicates the magnitude of current and the background color represents the valley polarization. The superposition of the incident and scattered waves produces characteristic constructive and destructive interference patterns, giving rise to the fine spatial structure of the current flow and polarization in the vicinity of the dot.
The two valleys exhibit a clear mirror symmetry: the $K'$ current distribution is the reflection of the $K$ distribution with respect to the $x$ axis, and the two valley contributions become identical at $\theta = 0$. This follows directly from the mode structure of the BLG Hamiltonian where time reversal symmetry maps $\tau \to -\tau$ and $m \to -m$ simultaneously, so each partial-wave channel of the $K$ valley has a counterpart in $K'$ with equal amplitude and opposite angular momentum, consistent with the valley-dependent scattering reported in Ref.~\cite{solomon2021valley}. Reversing the sign of $\Delta$ exchanges the scattering profiles of the two valleys: the angular distribution of $K$ under $\Delta > 0$ is identical to that of $K'$ under $\Delta < 0$, and vice versa. 

For experimental detection of valley-resolved currents, the extended interference pattern produced by a plane-wave incidence presents a major difficulty, since the scattered signal remains mixed with the incident beam over the entire system. Also, the valley dependent scattering shown in the far-field scattering (Fig.~\ref{fig:placeholder} (a)) is masked by interference effects and barely visible in the total current flow (Fig.~\ref{fig:placeholder} (b)). A natural way to overcome this issue is to use a narrow Gaussian beam, so that the valley-resolved scattered currents can be deflected outside the finite beam footprint, becoming spatially separated from the forward-propagating incident beam and therefore directly accessible to measurement. This motivates the use of Gaussian-beam excitation, for which we developed the corresponding theoretical framework in the preceding subsection and discuss associated results afterward.

\subsection{Gaussian incident beam}\label{Gaussian}

We construct the Gaussian beam shown in Fig.~\ref{fig:fig0}(b) using the real-space Dirac equation. By expressing the Dirac Hamiltonian in Eq.~\eqref{Hamiltonian} in Cartesian coordinates, with $k_x = k\cos\theta$, $k_y = k\sin\theta$, and using the correspondence $k_x \pm i k_y \rightarrow -i\,\partial_{\pm}$, with $\partial_\pm = \tau\partial_x \pm i\partial_y$, the Dirac equation $H\psi = E\psi$ can be rewritten in component form as follows: 
\begin{subequations}
\label{eq:four_first_order_cart}
\begin{align}
 -i\,\partial_- \psi_{b_2} + \tilde\gamma_1 \psi_{B_1} 
     &= \tilde{E}\psi_{A_2},\\
 -i\,\partial_+ \psi_{A_2} 
     &= \tilde{E}\psi_{b_2},\\
 -i\,\partial_- \psi_{B_1} 
     &= \tilde{E}\psi_{a_1},\\
 \tilde\gamma_1 \psi_{A_2} - i\,\partial_+ \psi_{a_1} 
     &= \tilde{E}\psi_{B_1},
\end{align}
\end{subequations}
which is similar to procedure used to derive the incident plane wave of the previous section, which is detailed in Ref.~\cite{solomon2021valley}.
Eliminating the auxiliary components $\psi_{b_2}$, $\psi_{a_1}$, and $\psi_{B_1}$ yields a fourth-order equation for $\psi_{A_2}$,
\begin{equation}
\left[
(\partial_x^2+\partial_y^2)^2
+2\tilde{E}^2(\partial_x^2+\partial_y^2)
+\tilde{E}^2(\tilde{E}^2-\tilde\gamma_1^2)
\right]\psi_{A_2}=0.\nonumber
\end{equation}

Assuming propagation along $+x$, we apply the paraxial ansatz $\psi_{A_2}(\mathbf{r}) = e^{ikx}f(\mathbf{r})$, where $f(\mathbf{r})$ is slowly varying on the scale $\lambda = 2\pi/k$. The wavevector $k$ satisfies the BLG dispersion relation
\begin{equation}
k^4 - 2\tilde{E}^2 |k|^2 + \tilde{E}^2(\tilde{E}^2 - \tilde\gamma_1^2) = 0,
\end{equation}
so that the constant terms in the fourth-order equation are exactly canceled. Retaining only leading order derivatives ($\partial_x^2 f \ll k\,\partial_x f$) then reduces the equation to the standard paraxial form
\begin{equation}
\left(\partial_y^2 + 2ik\,\partial_x\right) f(\mathbf{r}) = 0,
\end{equation}
which admits the Gaussian solution
\begin{equation}
f(x,y)=\frac{A_0}{\sqrt{x-\omega}}
\exp\!\left(\frac{iky^2}{2(x-\omega)}\right),
\end{equation}
in analogy with the well-known Gaussian solutions of the Helmholtz equation \cite{landesman1988gaussian}. A detailed study of this has been carried out in Ref. \cite{matulis2011application} for monolayer graphene.
Here, $\omega = u + iv$, where $u$ denotes the focal position along the $x$ direction and $v > 0$ determines the beam waist, $W_0 = \sqrt{2v/k}$. The angular divergence of the beam in the far field is correspondingly given by $\theta_{{beam}} = 2/(kW_0)$.

The remaining spinor components are recovered from Eqs.~\eqref{eq:four_first_order_cart} by substituting the paraxial $\psi_{A_2}$ and retaining leading order in $k$, yielding the four-component Gaussian beam
\begin{equation}
\psi^g(x,y)\simeq
\frac{A_0}{\sqrt{x - \omega}}
\exp\!\left(\frac{iky^2}{2(x-\omega)}\right)
\begin{pmatrix}
|\tilde{E}|\\
\eta_1\,\tau k\\
\eta_2\,\tau k\\
\eta_1\eta_2\,|\tilde{E}|
\end{pmatrix}
e^{ikx}.
\end{equation}
The Gaussian beam in the $x$–$y$ plane propagating along $x$, at energy $E = 0.05\,\gamma_1$, is shown in Fig.~\ref{fig:fig0}(b). Using $W_0 = 20\,l_0$ and $u = 0$ yields a beam divergence of $\theta_{{beam}} \approx 25^\circ$, as used in this work.

To incorporate this beam into the scattering framework, we project it onto cylindrical partial-wave modes,
\begin{equation} \label{expansion}
\psi_{\mathrm{inc}}^{g}(k,r)
=\sum_{m=-\infty}^\infty g_m\, \psi_{m,\mathrm{inc}}(k,r),
\end{equation}
where the projection coefficients are
\begin{equation}
g_m
= \frac{1}{\mathcal{N}_m}
\int \psi_{\mathrm{inc}}^{m,\dagger}(k,r)\,
\psi^g(x,y)\,
r\,dr\,d\theta,
\end{equation}
with $\mathcal{N}_m = \int |\psi_{\mathrm{inc}}^m|^2\, r\,dr\,d\theta$ 
the norm of each partial-wave mode. The scattering problem is then solved exactly as in the plane-wave case, with the Gaussian envelope encoded in $g_m$.

\begin{figure}
    \centering 
    \includegraphics[width=1\linewidth]{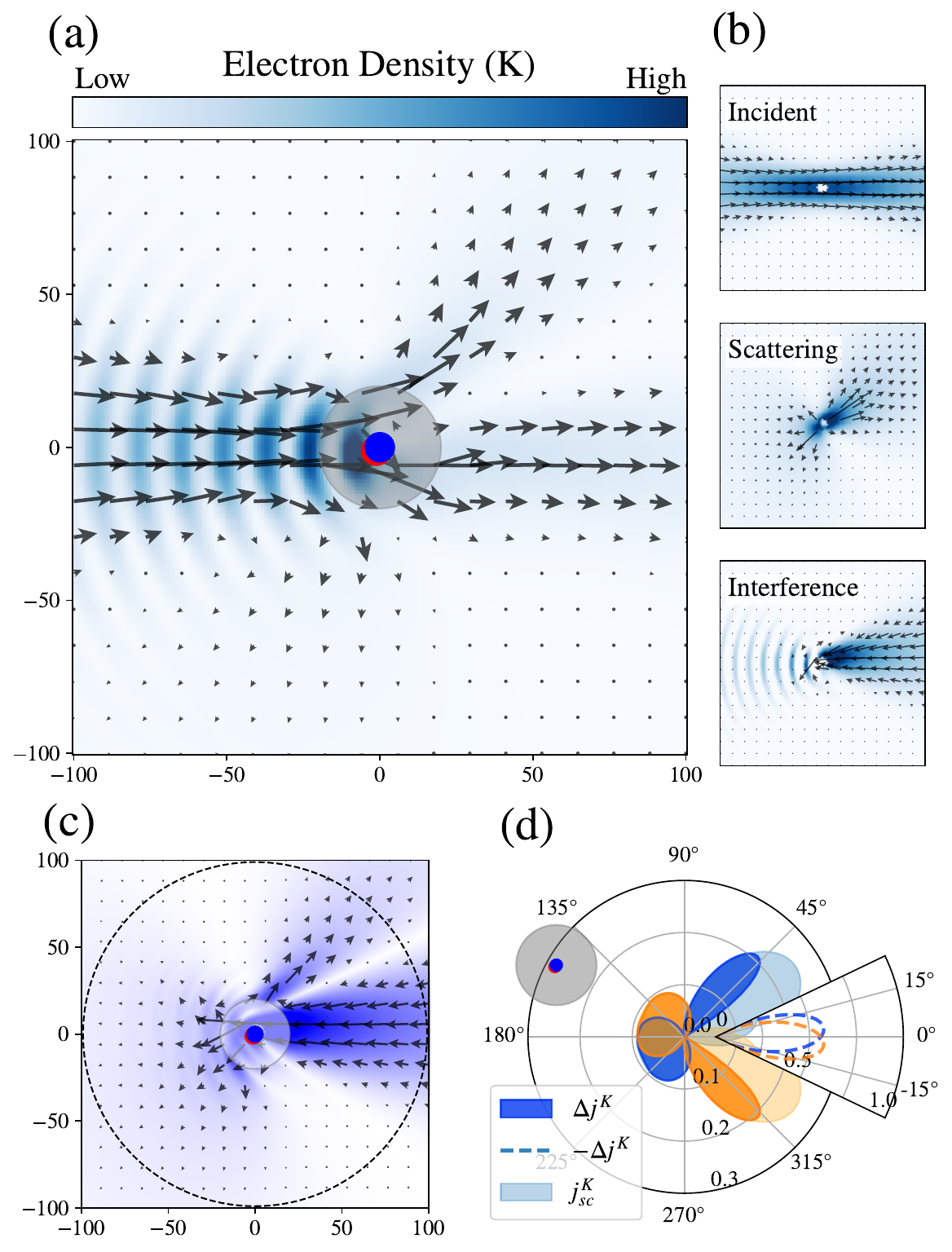}
    \caption{(a) Current flow (arrows) and density (color map) for the $K$ valley under Gaussian-beam incidence. (b) Decomposition into incident, scattered, and interference contributions. (c) Difference current map $\Delta j^K$. the $K'$ response is the mirror image of the $K$ response with respect to the $x$ axis for (a)-(c). (d) Angular dependence of $\Delta j^K$ at $r=100\,l_0$ denoted by dashed circle in (c). Light colors curves show far-field results; orange curves correspond to the $K'$ valley. The inset polar sector shows negative contributions within $\pm\theta_{{beam}}$, corresponding to forward-current suppression.}
    \label{fig:contributions}
\end{figure}


Once the coefficients $g_m$ are determined, the Gaussian incident spinor is reconstructed via Eq.~\eqref{expansion}. 
Imposing continuity of all four spinor components at $r = R$ as in Eq.~\eqref{continuity}, using the constructed Gaussian beam instead of a plane-wave incident state, then yields a linear system for the unknown scattering coefficients $c^{\mathrm{sc}}_{m,\pm}$ and $c^{\mathrm{tr}}_{m,\pm}$, exactly as in the plane-wave case.
Since the transverse profile of the Gaussian beam is fully encoded in the prefactors $g_m$, the remaining procedure after constructing the continuity equations, including extracting the coefficients, and computing physical observables, proceeds identically to the plane-wave treatment.

Fig.~\ref{fig:placeholder}(c) shows the far-field angular dependence of the valley-resolved and total scattered currents for a Gaussian beam incident on a single QD positioned at the center of beam. The angular profiles of the two valleys are clearly distinct and qualitatively consistent with the plane-wave results of Fig.~\ref{fig:placeholder}(a), confirming that the valley-dependent scattering is robust to the beam profile. The dashed curves show the angular dependence of the scattered current evaluated at the finite distance $r = 100\,l_0$ from the center for both valleys. Their close agreement with the far-field curves confirms that the far-field regime is well established at this distance.

Fig.~\ref{fig:placeholder}(d) presents the corresponding total current distribution and valley polarization map. Outside the beam footprint, the $K$ valley current is deflected upward and the $K'$ current downward, producing strong spatial valley separation. Within the beam, the forward current remains largely unpolarized due to the equal contribution of both valleys to the forward-scattered field at $\theta = 0$. 
Compared to plane-wave excitation shown in Fig.~\ref{fig:placeholder}(b), the Gaussian beam suppresses contributions from large $|m|$ channels away from the beam axis and consequently reduces interference of incident and scattered waves, allowing cleaner observation of valley-dependent deflected currents. 

Outside the narrow beam where deflected valley currents are well separated from the incident beam, the absence of the opposing valley's incident field enhances the numerator of Eq.~\eqref{polarization} while reducing the denominator, yielding significantly higher valley polarization values.


\subsection{Current decomposition}\label{decomposition}

To analyze in more detail the scattering of a Gaussian electron beam from a dot located at the beam center, we decompose the total wavefunction around the dot into incident and scattered contributions for each valley $\tau$:
\begin{equation}
\psi^\tau = \psi_{\mathrm{inc}}^\tau + \psi_{\mathrm{sc}}^\tau.
\end{equation}

Using the current equation $j^\tau = \psi^{\tau\dagger} 
\boldsymbol{\sigma}_\tau \psi^\tau$ naturally yields three contributions: 
the incident current $j_{\mathrm{inc}}^\tau = \psi_{\mathrm{inc}}^{\tau\dagger} 
\boldsymbol{\sigma}_\tau \psi_{\mathrm{inc}}^\tau$, the scattered current 
$j_{\mathrm{sc}}^\tau = \psi_{\mathrm{sc}}^{\tau\dagger} \boldsymbol{\sigma}_\tau 
\psi_{\mathrm{sc}}^\tau$, and the interference current

\begin{equation}
j_{\mathrm{int}}^\tau =
\psi_{\mathrm{inc}}^{\tau\dagger} \boldsymbol{\sigma}_\tau 
\psi_{\mathrm{sc}}^\tau
+
\psi_{\mathrm{sc}}^{\tau\dagger} 
\boldsymbol{\sigma}_\tau \psi_{\mathrm{inc}}^\tau,
\end{equation}
so that the total valley-dependent current is decomposed into all three contributions:
\begin{equation}
j^\tau = j_{\mathrm{inc}}^\tau + j_{\mathrm{sc}}^\tau + j_{\mathrm{int}}^\tau.
\end{equation}

The second term, $j_{\mathrm{sc}}^\tau$, approaches the far-field scattered current in the asymptotic regime (here at $R = 100\,l_0$, as shown in Fig.~\ref{fig:placeholder} (c)), and is commonly used to characterize scattering. However, this contribution alone does not capture the full current flow: its angular pattern does not necessarily reflect the spatial current distribution near the dot. The incident and interference contributions are therefore also required for a complete description. 
In the following, we discuss these contributions for the $K$ valley for the case shown in Fig.~\ref{fig:placeholder}(c) and (d). Fig.~\ref{fig:contributions}(a) shows the total $K$-valley current ($j^K$), which exhibits a deflection toward the upper side, consistent with Fig.~\ref{fig:placeholder}(d). The color shading indicates the electron density of the $K$ valley.

The three contributions to $j^K$ are shown separately in Fig.~\ref{fig:contributions}(b). The top panel shows the incident current $j_{\mathrm{inc}}$, which follows the Gaussian beam profile undisturbed, with higher electron density and larger current near the beam center, decaying toward the sides. The middle panel shows the purely scattered current $j_{\mathrm{sc}}$ from the dot, directed predominantly toward the upper-right side, and a weaker oppositely directed backscattered portion, consistent with the $K$-valley far-field scattering profile in Fig.~\ref{fig:placeholder}(c). As expected, the scattered current is considerably smaller in magnitude than the incident current.

The bottom panel of Fig.~\ref{fig:contributions}(b) shows the interference current $j_{\mathrm{int}}$, which is nonzero only in the spatial region where the incident and scattered waves overlap. On the right side of the dot, where both contributions are outgoing, the interference current is directed predominantly toward the dot and exceeds the scattered current in magnitude, making it the dominant modification to the total current in this region.
In other words, a reduction in incident current here requires a strong incoming current to cancel it, this can only come from interference term, as scattering is strictly outgoing.
In turn, a strong interference current relies on the scattering wavefunction, which in turn gives rise to a scattering current (smaller than the interference).
Analyzing only the scattering current (as in Fig.~\ref{fig:placeholder}(c)) misses this effect, and can imply strong forward scattering or enhancement of current in the forward direction, whereas actually the forward current is suppressed.

To isolate the effect of the dot on the measurable current, we consider a new metric: the difference between the total and incident currents, thereby retaining both scattering and interference effects:
\begin{equation}
\delta j^\tau \equiv j^\tau - j_{\mathrm{inc}}^\tau  = j_{\mathrm{sc}}^\tau  + 
j_{\mathrm{int}}^\tau, 
\quad
\Delta j^\tau \equiv \frac{\delta j^\tau}{\max(j^\tau_{\mathrm{inc}})},
\end{equation}
where $\Delta j^\tau$ is a normalized quantity defined with respect to the peak incident current, so that value of $\Delta j=-1$ corresponds to a total suppression of the incident beam.

This quantity is mapped for the $K$ valley in Fig.~\ref{fig:contributions}(c). A strong inward-directed current is visible on the right side of the dot, within the beam footprint, originating primarily from the interference term. A branch of outward current toward the upper-right lies outside the incident beam footprint, where interference is absent and the signal is therefore purely due to scattering. A weaker backward current from backscattering is also visible.

To analyze the angular profile, we evaluate $\Delta j^\tau$ at the distance of $r = 100\,l_0$ from the center of the beam, which is indicated by the dashed circle in Fig.~\ref{fig:contributions}(c). This distance lies well within the far-field regime, consistent with the agreement shown by the dashed curves in Fig.~\ref{fig:placeholder}(b) for scattered current. 
The main polar plot in Fig.~\ref{fig:contributions}(d) corresponds to positive values of $\Delta j^\tau$, indicating regions of net current enhancement around the dot. The inset polar plot highlights the negative contributions within the angular range of the incident beam footprint, $\pm\theta_{{beam}}$, located in front of the dot. These negative values reflect the suppression of forward current by destructive interference.

The solid dark-blue curve corresponds to the $K$ valley. The strong positive lobe on the upper side coincides with the dominant scattering direction (light-blue curve), confirming that scattering part governs the signal outside the beam, while the rest of scattered contribution lies inside the beam and is influenced by large interference current as shown in inset. The negative values in the forward direction (dashed blue curve) quantify the fraction of the incident forward current suppressed by interference. The corresponding results for the $K'$ valley are shown in orange; which is the reflection of the $K$ profile across the $x$-axis. 
We therefore conclude that qualitative current flow is more clearly captured by $\Delta j$ than by $j_{\mathrm{sc}}$ alone.

\subsection{Electron multiple scattering in a Gaussian beam}\label{multiple scattering}

We now extend the approach to an array of $N$ circular gated QDs in BLG, where dot $\ell$ has radius $R_\ell$ and layer potentials $(V_{1,\ell}, V_{2,\ell})$. The center of dot $\ell$ is located at $\mathbf{b}_\ell = (b_\ell, \gamma_\ell)$ in the global coordinate system defined by the incident beam axis. A point $\mathbf{r}$ in the global frame corresponds to $\mathbf{r}_\ell = \mathbf{r} - \mathbf{b}_\ell = (r_\ell, \theta_\ell)$ in the local frame of dot $\ell$; all wave functions are subsequently expressed in the local frame of the relevant dot;    similar to relevant works on monolayer graphene \cite{ildarabadi2025tunable,sadrara2019dirac}.

The incident Gaussian beam, re-expanded around dot $\ell$, is
\begin{equation}
\psi_{\mathrm{inc}}^{g}(k, r_\ell)
=
\sum_{m=-\infty}^{\infty}
g_m^{(\ell)}\, \psi_{\mathrm{inc}}^m(k, r_\ell),
\end{equation}
with projection coefficients
\begin{equation}
g_m^{(\ell)}
=
\frac{1}{\mathcal{N}_m}
\int
\psi_{\mathrm{inc}}^{m,\dagger}(k, r_\ell)\,
\psi^{g}(x_\ell, y_\ell)\,
r_\ell\, dr_\ell\, d\theta_\ell,
\end{equation}
where $x_\ell = r_\ell\cos\theta_\ell$, $y_\ell = r_\ell\sin\theta_\ell$, 
and $\mathcal{N}_m$ is the norm of the $m$-th partial-wave mode.

The total wave field outside all dots is decomposed as the incident beam plus the scattered one.
The total scattered wave is written as the sum of the contribution from dot $\ell$ and the contributions from all other dots,
\begin{equation}
\psi_{\mathrm{sc}}(k,r)
=
\psi_{\mathrm{sc},\ell}(k,r_\ell)
+
\sum_{j\neq \ell}^{N}
\psi_{\mathrm{sc,j}}(k,r_j).
\end{equation}
where each contribution takes the form
\begin{align}
\psi_{\mathrm{sc},j}(k, r_j)
&=
\frac{1}{\sqrt{2(\tilde{E}^2+|k|^2)}}
\sum_{m=-\infty}^{\infty}
c^{\mathrm{sc,j}}_{m,k}\, i^m \nonumber\\
&\quad\times
\psi^{0,H}_m(k r_j, \theta_j).
\end{align}

To impose the boundary condition at dot $\ell$, we re-express the scattered field from dot $j \neq \ell$ in the local frame of dot $\ell$. Defining the relative displacement $\mathbf{b}_{\ell j} = \mathbf{b}_\ell - \mathbf{b}_j = (b_{\ell j}, \beta_{\ell j})$, Graf's addition theorem gives, for $r_\ell < b_{\ell j}$,
\begin{align}
H^{(1)}_m(k r_j)\,e^{im\theta_j}
&=
\sum_{n=-\infty}^{\infty}
H^{(1)}_{m-n}(k b_{\ell j})\,
e^{i(m-n)\beta_{\ell j}} \nonumber\\
&\quad\times
J_n(k r_\ell)\,e^{in\theta_\ell}.
\end{align}
Applying this identity to each spinor component, the contribution from dot $j$ in the frame of dot $\ell$ becomes
\begin{equation}
\psi_{\mathrm{sc},j}(k, r_\ell)
=
\sum_{m,n}
c^{\mathrm{sc,j}}_{n,k}\,
T^{(\ell\leftarrow j)}_{m,n}\,
i^{m}\,\psi^{0,J}_{m}(k r_\ell, \theta_\ell),
\end{equation}
where the transfer matrix elements are
\begin{equation}
T^{(\ell\leftarrow j)}_{m,n}
=
\frac{i^{n-m}}{\sqrt{2(\tilde{E}^2+|k|^2)}}\,
H^{(1)}_{n-m}(k b_{\ell j})\,
e^{i(n-m)\beta_{\ell j}}.
\end{equation}
Inside dot $\ell$, the transmitted wave is
\begin{equation}
\psi_{\mathrm{tr},\ell}(r_\ell, \theta_\ell)
=
\sum_{m=-\infty}^{\infty}
c^{tr,\ell}_{m,k}\, i^{m}\,
\psi^{V_\ell}_{m}(k_{\ell} r_\ell, \theta_\ell).
\end{equation}
Imposing continuity of all four spinor components at $r_\ell = R_\ell$ gives
\begin{align}
\psi^g_{\mathrm{inc}}(k_-^0,R_\ell)
&+
\psi_{\mathrm{sc},\ell}(k_-^0,R_\ell)
+
\psi_{\mathrm{sc},\ell}(k_+^0,R_\ell)
\nonumber\\
&+
\sum_{j\neq \ell}
\Big[
\psi_{\mathrm{sc},j}(k_-^0,R_\ell)
+
\psi_{\mathrm{sc},j}(k_+^0,R_\ell)
\Big]
\nonumber\\
&=
\psi_{\mathrm{tr}}(k_-,R_\ell)
+
\psi_{\mathrm{tr}}(k_+,R_\ell).
\end{align}

Projecting onto angular momentum channel $m$ yields four coupled equations per dot. Solving the first two components for $c^{tr,\ell}_{m,k}$ and substituting into the remaining two gives a reduced linear system of dimension $2MN$ for the scattering coefficients $c^{\mathrm{sc,\ell}}_{m,k}$, where $M$ is the number of retained angular modes. This system is solved numerically, after which all wave functions and physical observables follow directly from the coefficients.

\section{Results}\label{Results}
We present results for different classes of multi-dot architectures, each built from the single dot scattering response established in Sec.~\ref{decomposition} where collective interference between multiple dots inevitably modifies the resulting transport properties. We first consider simple arrays of identical dots, which augment valley splitting through collective interference, then turn to oppositely gated pairs, which act as valley filters, and finally combine these elements into a multi-component architecture for directional steering of a highly valley-polarized current. Throughout, we fix  $E = 0.05\,\gamma_1$, $R = 5\,l_0$, $|\Delta| = 0.4\,\gamma_1$, and $W_0 = 20\,l_0$, and vary only the dot arrangement and gating configuration.

\subsection{Simple identical dot architectures}
\begin{figure}
    \centering
    \includegraphics[width=1\linewidth]{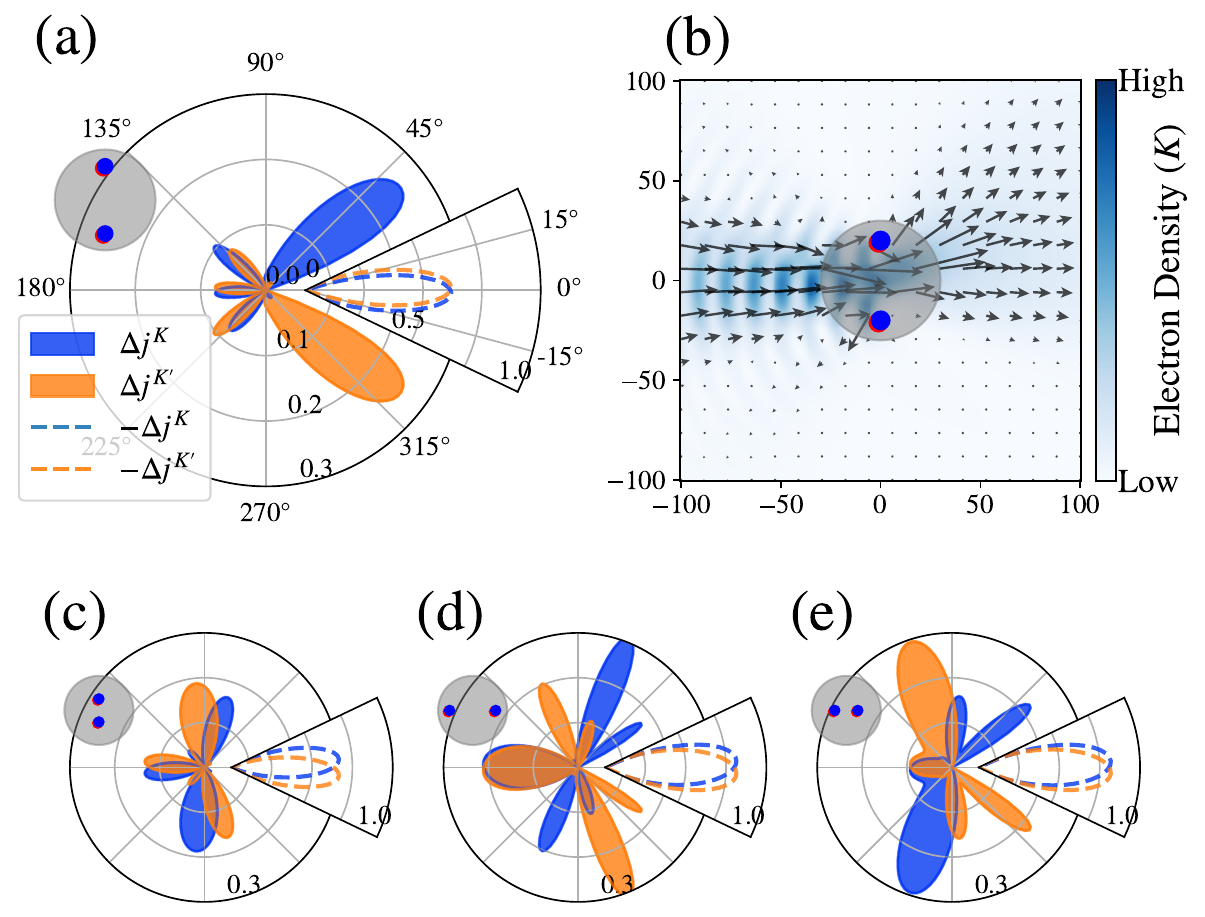}
    \caption{
    Valley-resolved current distributions resulting from multiple scattering by two identical QDs under Gaussian-beam incidence. 
    (a) $\Delta j$ for vertically aligned dots with $d=8R$ (upper-left inset), showing efficient valley splitting. Blue and orange curves correspond to the $K$ and $K'$ valleys, respectively; the dashed curves in the right inset show the negative part of $\Delta j$ in the forward-direction beam region. 
    (b) Corresponding $K$-valley current-flow (arrows) and electron density (color map); the $K'$ response is the mirror image of the $K$ response with respect to the $x$ axis.
    (c)–(e) $\Delta j$ for vertically aligned dots with $d=4R$, and horizontally aligned dots with $d=8R$ and $4R$, respectively. 
    }
    \label{fig:splitter}
\end{figure}


For a system of two identical dots, the overall scattering pattern remains qualitatively similar to that of a single dot, with modifications that we discuss below. We focus on the $K$ valley; the $K'$ valley exhibits the corresponding behavior mirrored with respect to the $x$-axis, since the system is symmetric about this axis.
We first consider two dots separated vertically by a center-to-center distance $d=8R$, as shown in Fig.~\ref{fig:splitter}(a). In this configuration, the angular positions of the dominant valley-resolved lobes are largely preserved compared to single dot (Fig.~\ref{fig:contributions}(d)), and only minor modifications to the current distribution are observed. This behavior can be understood from the corresponding current map for the $K$ valley shown in Fig.~\ref{fig:splitter}(b).

In this case, the two dots are positioned near the edges of the Gaussian beam waist, where the beam intensity is lower. In particular, there is very small part of incident current above the upper dot and below the lower dot. As a result, the portion of the $K$-valley beam that would normally be deflected upward by upper dot (as in the single dot case in Fig.~\ref{fig:contributions}(a)) is largely suppressed. Instead, because the dots are sufficiently separated, the beam propagates through the region between them and is deflected upward primarily by the lower dot. Since this deflection originates from the lower dot, the corresponding lobe is slightly shifted toward smaller angles compared to the single dot case.

At the same time, the part of the K-valley beam that propagates straight through the system, passing between the two dots, is mainly associated with the upper dot, in analogy with the portion of the beam passing below a single dot in Fig.~\ref{fig:contributions}(a).
The backscattered components are also suppressed in this configuration, again due to the reduced beam intensity at the dot positions near the beam edges. 

When the two dots are brought closer together to $d=4R$ as in Fig.~\ref{fig:splitter} (c), they are positioned more toward the central region of the beam. A finite portion of the beam now exists above the upper dot and below the lower dot. As a result, the upper dot is directly illuminated by part of the $K$-valley beam, leading to an upward deflection similar to the single dot case in Fig.~\ref{fig:contributions}(a), though with slightly reduced magnitude due to the decay of beam intensity away from the center. This contribution appears as the blue lobe in the upper half-plane of Fig.~\ref{fig:splitter}(c), directed more transversely due to the off-center position of the dot.
Because the dots are now in close proximity, the beam that would otherwise be deflected upward by the lower dot is partially blocked by the upper dot. This contribution is consequently redirected into backscattering, combining with the intrinsic backscattering from both dots to produce the pronounced lobe in the lower half-plane directed more toward the downward transverse direction.

Overall, closer vertical spacing enhances transverse deflection through multiple scattering; however, this comes at the cost of increased overlap between the $K$ and $K'$ contributions, thereby reducing valley selectivity. The angular positions of the scattering lobes can therefore be controlled via the vertical separation, which shifts the maxima and minima of the angular distribution and thus tunes the effective deflection angles. A residual portion of the $K$-valley beam still transmits through the gap or passes beneath the lower dot with weak deflection.

If the identical dots are aligned horizontally and are sufficiently separated ($d=8R$), as shown in Fig.~\ref{fig:splitter}(d), the overall scattering can be understood as a sequential combination of two single dot responses. The first dot plays the dominant role, as it is directly illuminated by the incident beam, while only a reduced portion of the beam reaches the second dot. The second dot therefore introduces a weaker modification. This is reflected in the angular profile, where the larger blue lobe in the upper half-plane originates from the first dot and the smaller lobe from the second dot.

When the horizontally separated dots are brought closer to $d=4R$, as in Fig.~\ref{fig:splitter}(e), interference effects become more significant. The beam deflected by the first dot immediately re-encounters the second dot, leading to two distinct lobes in the upper half-plane. The backscattered component is also enhanced by the combined action of both dots, producing a pronounced lobe in the lower half-plane.

Although the overall improvement over the single-dot configuration remains limited for horizontal alignment, the two-dot analysis reveals that the vertical and horizontal separations play distinct roles in shaping the scattering pattern: the vertical separation primarily controls the angular deflection of the scattered beams, whereas the horizontal separation mainly modifies the relative strengths of the two scattering lobes. These insights point to the key mechanisms underlying more effective multi-dot designs: the geometric arrangement controls both the angular distribution and the relative strength of different scattering channels, motivating the more complex architectures considered in the following section.

\begin{figure}
    \centering
    \includegraphics[width=1\linewidth]{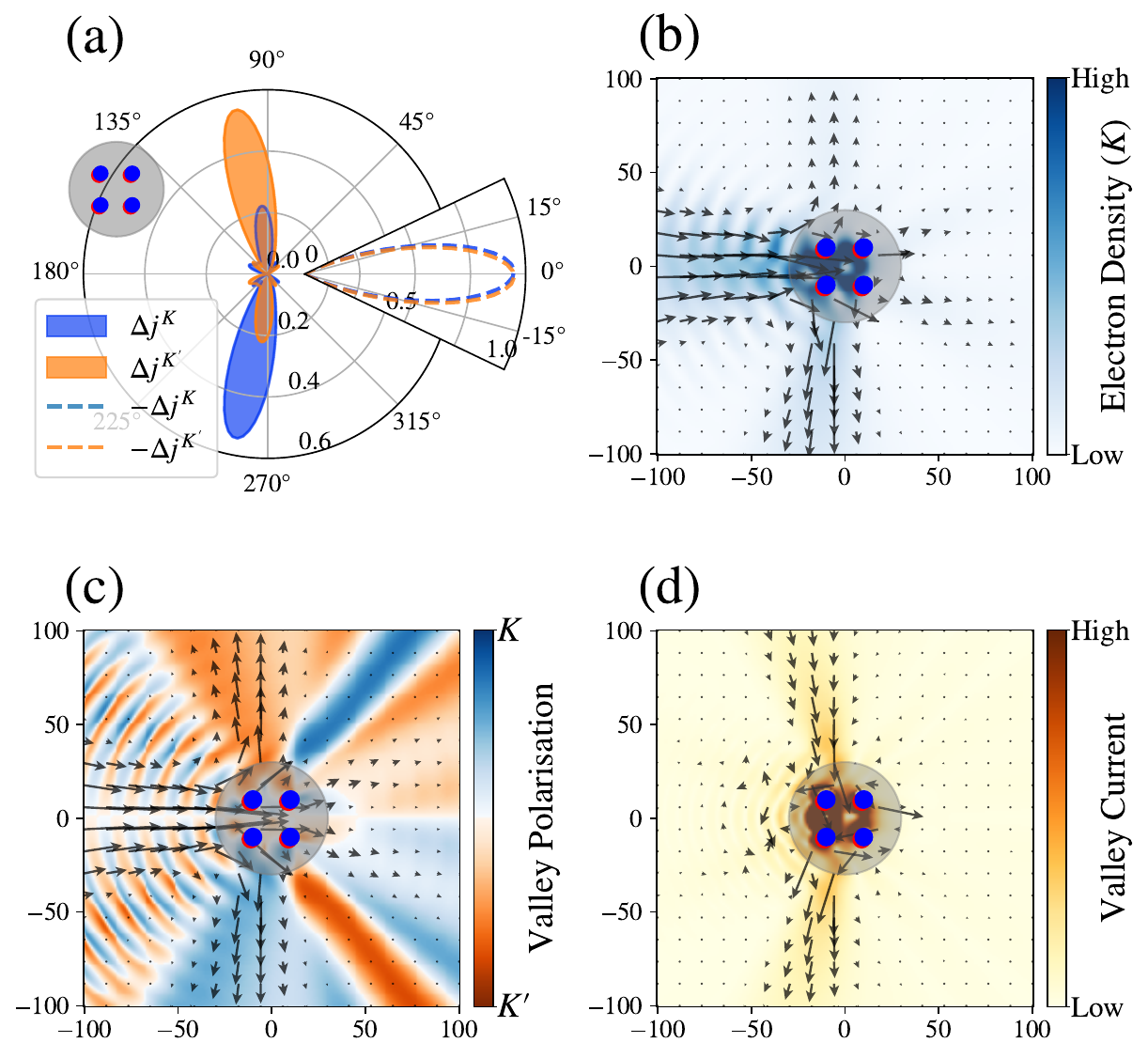}
    \caption{Valley-resolved current for a $2\times2$ array of identical QDs with $d=4R$. (a) $\Delta j$, showing strong transverse deflection and reduced valley overlap, forming an efficient valley splitter. (b) Corresponding $K$-valley current-flow (arrows) and electron density (color map); the $K'$ response is its mirror image with respect to the $x$-axis. (c) Total current map overlaid on the valley polarization color map. (d) Valley current map, showing strong transverse valley separation.}
    \label{fig:splitter2}
\end{figure}

By combining the mechanisms discussed above in multi-dot configurations, one can engineer structures with tailored scattering responses. In particular, a $2\times2$ array, Fig.~\ref{fig:splitter2}, exhibits strong transverse deflection with minimal overlap between the $K$ and $K'$ contributions, together with substantial suppression of the forward beam. The valley-resolved $\Delta j$ at $r=100\,l_0$ in Fig.~\ref{fig:splitter2}(a) clearly demonstrates this behavior, with the dashed curve indicating strongly reduced forward transmission.

The current maps further highlight the valley separation: the $K$ valley current in Fig.~\ref{fig:splitter2}(b) is strongly deflected in the downward transverse direction, while the $K'$ contribution (not shown) appears as its mirror image with respect to the $x$-axis. Since the deflected current lies well outside the beam footprint, it originates predominantly from scattering, with negligible interference from the incident beam.
This configuration thus realizes an efficient valley splitter, where both directionality and selectivity are optimized. The transverse regions exhibit strong current with high valley polarization, as shown in Fig.~\ref{fig:splitter2}(c), and the valley current $j_V = j^K - j^{K'}$ in Fig.~\ref{fig:splitter2}(d) provides a direct real-space signature of transverse valley flow.

The strength of this effect is quantified by the valley Hall angle, which characterizes the efficiency of converting a longitudinal charge current into a transverse valley current. It is defined as $\tan \theta_{\mathrm{VH}} = j_V^{y} / j_{\mathrm{inc}}^{x}$, where $j_{\mathrm{inc}}^{x} = j_{\mathrm{inc}}^{K} + j_{\mathrm{inc}}^{K'}$, and $x$ and $y$ denote the corresponding current components integrated over the full cross-section. Evaluating the currents on a circular contour at a distance $r = 100\,l_0$ around the dot system, we obtain $\tan \theta_{\mathrm{VH}} \simeq 0.3$, indicating a pronounced transverse response. This value is of the same order of magnitude as those reported for tilt-induced tunneling \cite{zhang2023tunneling}, electrostatic barrier scattering \cite{zeng2024tunneling}, and skew scattering from impurities \cite{xu2017geometric}.
This places the system in a regime of efficient valley-current generation, highlighting the effectiveness of valley-contrasting electron optics for producing measurable transverse valley currents.

\subsection{Oppositely gated dot architectures}

\begin{figure}
    \centering
    \includegraphics[width=1\linewidth]{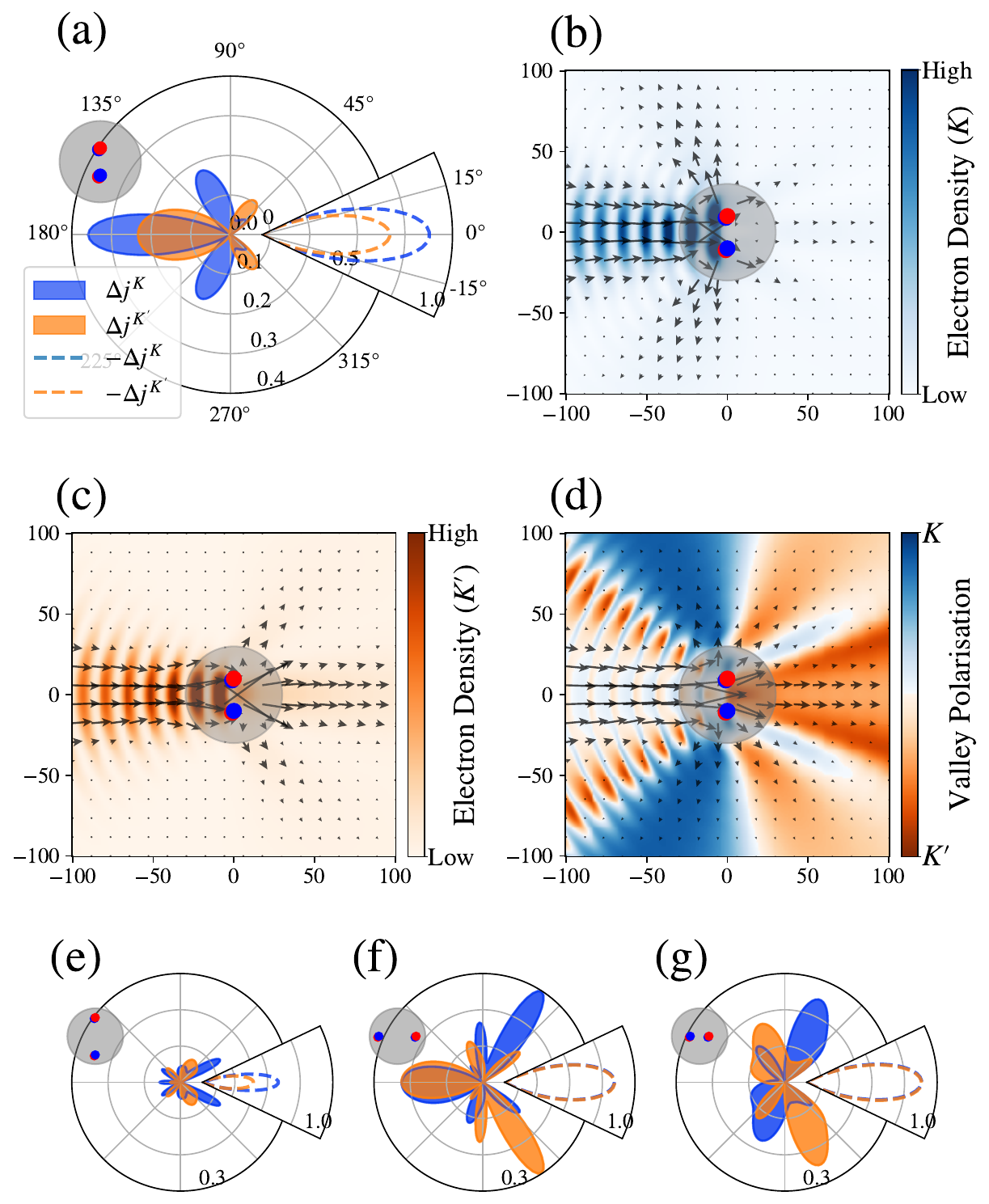}
    \caption{Valley-resolved current for two oppositely gated QDs. (a) $\Delta j$ for vertically aligned dots with $d=4R$, showing a pronounced valley-filtering effect. (b) and (c) Current-flow (arrows) and electron density (color map) for the $K$ and $K'$ valleys, respectively. (d) Total current flow (arrows) and valley polarization (color map).  (e) $\Delta j$ for vertically aligned dots with $d=8R$. (f) and (g) $\Delta j$ for horizontally aligned dots with $d=8R$ and $4R$, respectively.}
    \label{fig:filter}
\end{figure}

While systems of identical dots produce a mirror-symmetric response between the two valleys, introducing oppositely gated dots breaks this symmetry because the two gating configurations affect the $K$ and $K'$ valleys in opposite ways. In the single dot case of Fig.~\ref{fig:contributions}, the response observed for the $K$ valley is reversed when the layer potentials are swapped, i.e., a dot with $+\Delta$ deflects the $K$ 
valley upward and $K'$ downward, while a dot with $-\Delta$ does the 
opposite. Combining such oppositely gated dots, therefore allows the response of 
each valley to be engineered independently, rather than simply mirrored; whereas identical dots primarily generate a splitting effect, oppositely gated dots can produce a filtering effect by enhancing the transmission of one valley while suppressing the other.

Fig.~\ref{fig:filter}(a)-(d) shows a system with two oppositely gated dots separated vertically by $d=4R$. For the $K$ valley, the upper dot would normally deflect the beam downward while the lower dot deflects it upward; with vertical separation $d=4R$, these two deflection paths are mutually blocked by the opposing dot, so the $K$-valley beam is predominantly backscattered. For the $K'$ valley, by contrast, the reversed scattering profile of each dot causes both dots to deflect the beam along trajectories that pass through the gap between them. While a substantial fraction of the $K'$ current is still backscattered, a significant portion is transmitted forward through the gap. The $K$ current, in contrast, is much more strongly suppressed in the forward direction, producing a pronounced contrast between the two valleys and thereby realizing a valley-filtering effect.
As seen in the inset of Fig.~\ref{fig:filter}(a), nearly $90\%$ of the $K$-valley forward current is suppressed, while panels (b) and (c) show the current maps corresponding predominantly to backscattered $K$ and transmitted $K'$ valleys, respectively, as further evidenced by the total current and valley polarization shown in panel (d).

The angular profiles of $\Delta j$ for larger separations and different alignments of two oppositely gated dots are shown in Fig.~\ref{fig:filter}(e)-(g). Increasing the vertical separation causes the two dots to act more independently: the lower dot scatters part of the $K$ beam upward and part of the $K'$ beam downward, while the upper dot produces the opposite effect. The larger separation also allows a greater portion of both valleys to pass through the gap, though still with an asymmetry similar to that in panel (a). As a result, the filtering effect is weakened, as reflected by the reduced negative values of $\Delta j$ in the inset of Fig.~\ref{fig:filter}(e).

Horizontally aligned oppositely gated dots do not produce valley-filtering effect. Because the two dots are along the propagation direction, the overall response can be viewed as a sequential scattering process in which each dot produces a valley profile that is the mirror image of the other with respect to the $x$ axis. Consequently, the final current distribution also preserves this mirror symmetry between valleys, so horizontally aligned opposite dots exhibit a more valley-splitting character than a filtering response. The first dot plays the dominant role, partially blocking the beam before it reaches the second dot, as shown in Figs.~\ref{fig:filter}(f) and (g). The larger upward-deflected $K$ lobe and downward-deflected $K'$ lobe are mainly associated with the first dot, while the smaller lobes in the opposite directions originate from the second dot. For smaller horizontal separations (Fig.~\ref{fig:filter}(g)), this screening effect becomes stronger due to increased blocking by the first dot.


\subsection{Multi-component valley-current engineering}

\begin{figure}
    \centering
    \includegraphics[width=1\linewidth]{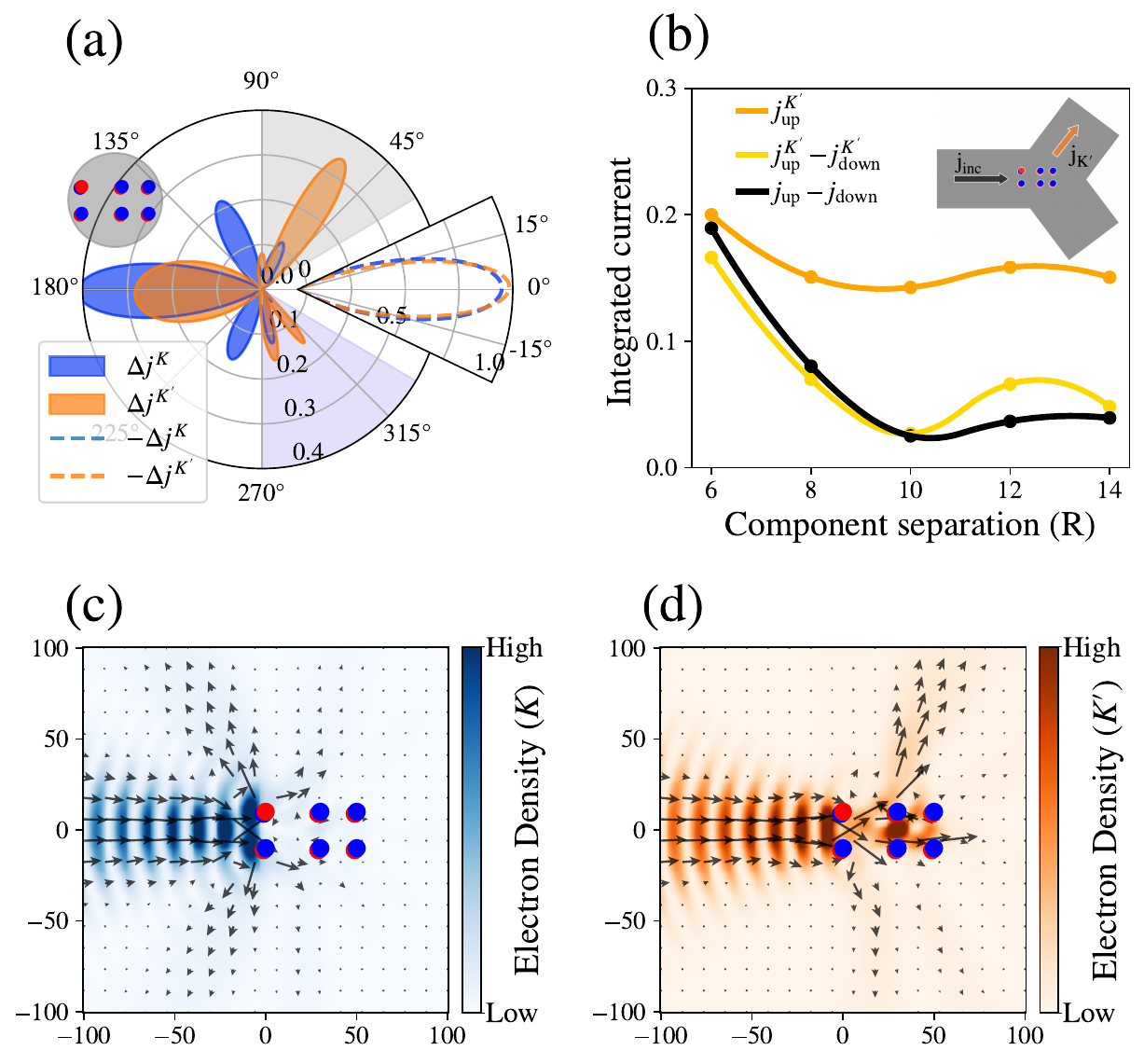}
    \caption{Valley-current engineering in a multicomponent architecture composed of a four-dot splitter followed by a two-dot filter. (a) $\Delta j$ for a component separation of $8R$, showing strong upward deflection of the $K'$ current and suppressed forward transmission (dashed inset curves). (b) Integrated current asymmetry between the upper and lower angular regions in panel (a) versus component separation. The black curve shows the measurable total-current difference in the Hall-bar setup schematically illustrated in the inset. (c),(d) Current-flow (arrows) and electron density (color map) for the $K$ and $K'$ valleys corresponding to panel (a).}
    \label{fig:multicomponent}
\end{figure}

Beyond the valley splitting and filtering effects discussed in the previous sections, understanding the mechanisms of electron multiple scattering in multi-dot architectures enables the design of combined functional components for valley-resolved beam engineering, including directional steering and filtering for practical device applications. Such multi-component architectures provide versatile on-demand control over valley-dependent current flow.

In Fig.~\ref{fig:multicomponent} we present a representative engineered platform consisting of a two-dot filter (as in Fig.~\ref{fig:filter}(a)) followed by a four-dot splitter (as in Fig.~\ref{fig:splitter2}), with a center-to-center separation of $8R$ between the two components.  The filter strongly backscatters the $K$ valley while transmitting the $K'$ component, which is subsequently redirected by the splitter toward the upper transverse direction. This combined structure generates a strongly valley-polarized current deflected toward the upper transverse direction, consisting predominantly of $K'$ carriers, while the current in downwards direction remains weak. At the same time, the forward current is nearly completely suppressed for both valleys, as indicated by the dashed curves in the inset of Fig.~\ref{fig:multicomponent}(a).

To quantify this further, we integrate the current over two opposite angular regions outside the Gaussian-beam footprint, $[\theta_{beam},\pi/2]$ and $[-\pi/2,-\theta_{beam}]$, indicated by the gray and purple shaded regions in Fig.~\ref{fig:multicomponent}(a). The resulting response provides an experimentally accessible signature of a highly valley-polarized current, with an almost pure $K'$ contribution in the upper angular region and only a weak residual current in the opposite direction. Such a signal could be measured using a Hall-bar geometry or a multi-probe transport setup, schematically illustrated in the inset of Fig.~\ref{fig:multicomponent}(b), where probes positioned over the corresponding angular regions at $r=100\,l_0$ collect the angular transverse currents.

To demonstrate the robustness of the effect, black curve in Fig.~\ref{fig:multicomponent}(b) shows the integrated current asymmetry between the upper and lower angular regions, $j(\mathrm{up})-j(\mathrm{down})$, as a directly measurable quantity plotted as a function of the center-to-center separation between the two components. Although the asymmetry decreases gradually with increasing separation, the effect remains pronounced throughout the range considered. The upper angular region continues to carry a strong $K'$ current (dark orange), while the yellow curve, representing the difference between the $K'$ currents in the upper and lower regions, closely follows the total asymmetry. This confirms that the measurable signal is dominated by the $K'$ contribution, while the $K$ current remains strongly suppressed for all separations considered. The gradual reduction of the overall effect with increasing separation mainly originates from the growth of the $K'$ current in the lower angular region, whereas the $K'$ current collected in the upper region changes comparatively little.
Current maps for the $K$ and $K'$ valleys corresponding to the configuration in Fig.~\ref{fig:multicomponent}(a), shown in Figs.~\ref{fig:multicomponent}(c) and \ref{fig:multicomponent}(d), respectively, provide a real-space visualization of the combined splitter-filter response. For the $K$ valley, the filter strongly suppresses transmission, leaving only a weak residual current beyond the structure. For the $K'$ valley, the current is transmitted through the filter and subsequently deflected by the splitter toward the upper transverse direction, in agreement with the far-field scattering profiles.

\section{Discussion and Conclusion}\label{Conclusion}
From a realization perspective, one approach to implementing the dot configurations considered here is dual-gate stacks, which employ independent local top and bottom gates, or alternatively an STM tip \cite{ge2021imaging,ge2020visualization,lee2016imaging}. These methods provide in-situ control over both local doping and layer asymmetry, which are central to our proposal, and have already been used to confine carriers in BLG QDs~\cite{banszerus2018gate,garreis2024long,jing2022gate,ge2020visualization}. This route enables a single platform that can be reconfigured between splitter, filter, and combined functionalities simply by tuning gate voltages. However, it remains challenging to scale to large arrays. As an alternative, dielectric patterning~\cite{forsythe2018band,barcons2022engineering} offers a more scalable route, where engineered spatial variations in dielectric thickness or permittivity create fixed local potentials using a smaller number of gates, at the expense of reduced dynamic tunability.

The parameters used throughout this work correspond to a Fermi energy of approximately $19~\mathrm{meV}$, a dot radius of $10~\mathrm{nm}$, and a dot layer asymmetry of $152~\mathrm{meV}$. These parameters are experimentally realistic for dual-gated BLG devices~\cite{zhang2009direct,ohta2006controlling}, and dot sizes of this scale are consistent with electrostatically defined structures reported in recent experiments~\cite{banszerus2018gate,ge2020visualization}, supporting the ballistic transport regime assumed here.

The scattering behavior is governed by the dimensionless parameter $kR$. Since BLG has a quadratic low-energy dispersion, $E \propto |k|^2$, the same scattering response is obtained under the scaling $R \to \alpha R$ and $E \to E/\alpha^2$, provided the incident beam size is rescaled by the same factor $\alpha$. This scaling invariance shows that the platform is not tied to a specific device size, but can be implemented across a wide range of experimentally accessible geometries.

To conlude, we developed a theoretical framework for valley-dependent electron optics in BLG based on electrostatically defined quantum-dot structures, using a four-band continuum Hamiltonian for AB-stacked BLG and a generalized Mie-type multiple-scattering formalism. Starting from the valley-contrasting scattering of a single gated dot under plane-wave incidence, we extended the analysis to Gaussian beams, which provide a more realistic injection profile. In contrast to plane-waves, the Gaussian beam suppresses extended interference via spatial separation between the incident beam and deflected valley currents. Decomposition of the current shows that scattering dominates outside the beam footprint, while interference governs the response within it.

The Gaussian beam is incorporated through projection onto partial-wave modes, allowing direct use of the same framework while retaining full angular resolution. A single dot produces a valley-polarized deflected current, establishing the basic mechanism for valley separation. In multi-dot systems, interference and geometry enable further control: identical dots lead to valley splitting, while oppositely gated dots act as valley filters. The effects of alignment, position within the beam, and inter dot spacing were systematically analyzed.

A $2\times2$ array of identical dots generates a strongly transverse, valley-polarized current with suppressed forward transmission, forming an efficient splitter, whereas two oppositely gated dots realize efficient valley filtering. Combining these elements produces a directionally steered, highly valley-polarized current with near-complete suppression of forward flow, providing a practical route for Hall-bar detection. These results demonstrate that engineered dot arrays enable controllable shaping of valley-resolved currents using purely electrostatic means, offering a scalable platform for valleytronic devices.

\begin{acknowledgements}
We acknowledge funding from the Irish Research Council under the Government of Ireland Postdoctoral Fellowship Program and from Research Ireland under the Frontiers for the Future program (24/FFP-P/12941). We also acknowledge the use of the gazelle computational facility in the School of Physical Sciences at DCU, which is supported by Intel Ireland.

\end{acknowledgements}


\end{document}